# Scalable Role-based Access Control Using The EOS Blockchain

Mohsin Ur Rahman

Department of Computer Science, University of Pisa
mohsinur.rahman@di.unipi.it
University of Pisa

**Abstract.** Role-based access control (RBAC) policies represent the rights of subjects in terms of roles to access resources. This research proposes a scalable, flexible and auditable RBAC system using the EOS blockchain platform to meet the security requirements of organizations. The EOS blockchain platform for developing smart contract and decentralized applications (DAPPs) aims to address the scalability problem found in existing blockchain platforms such as Ethereum and Bitcoin. This smart contract platform aims to eliminate transaction fees while conducting millions of transactions per second. In our proposed approach, the EOS blockchain transparently stores RBAC policies. Administrative roles control access to resources at a higher level according to the way organizations perform business. An organization creates roles, role hierarchies and constraints to regulate users' actions. Therefore, once an RBAC framework is established, the administrative user (issuer) only needs to grant and revoke roles to support changes in the organizational struc- ture. Moreover, resource owners delegate access rights to other subjects (in the same or different organization). Our proposed blockchain-based RBAC supports delegation capabilities using gaseless transactions which makes it adoptable and appealing in a large number of application sce- narios. Our proposed solution is application-agnostic and well-suited for diverse use cases. Existing state-of-the art security frameworks are not suitable due to the difficulty of scale, higher cost and single point of fail- ure. Consequently, organizations demand a scalable, cost-effective and lightweight access control solution which can better protect their pri- vacy as well. A proof of concept implementation is developed based on the EOS blockchain. In particular, we deploy our proposed RBAC smart contract on the EOS Kylin testnet. Our experimental results and analysis clearly show that our EOS blockchain-based RBAC outperforms exist- ing blockchain platforms in terms of cost, latency, block generation time, contract execution time and throughput.

Blockchain, Access Control, Scalability, Security, RBAC, Smart Contract

## 1 Introduction

From a functional viewpoint, the Role-based access control (RBAC) concept is that operations showing actions are associated with roles and subjects are granted suitable roles according to the organization's requirements. A role can be simply defined as the responsibilities assigned to a subject working in an organization. RBAC uses a many-to-many relationships between the subjects, roles and operations [1]. Many to many relationship means that a subject can be assigned many roles, and a single role can have several subjects. Similarly, a role can be assigned many permissions and the same permission can be assigned to many roles. The operations that are associated with roles constrain members of the role to a specified set of actions. RBAC supports a wide range of operations depending on the type of system such as read, write, insert, delete, update and so on. RBAC policies can be enforced when a subject attempts to execute an operation on an object, at the time when subjects are authorized as members of a role and during role activation (e.g., when a role is established as part of a user' active session).

The modern concept of RBAC consists of roles, role hierarchies and constraints on subject-role relationship and role-permission relationship. The role hierarchy defines an inheritance relationship among roles, that is, that one (junior) role may implicitly include the operations, constraints, and objects that are associated with another (senior) role. Role hierarchies simplify security administration and maintenance because junior roles implicitly include the permissions associated with senior roles [2]. RBAC supports constraints on almost all parts including permissions, roles and assignment relations to provide greater flexibility [1] [3] . The administrative user (issuer) can put a variety of constraints including cardinality constraints and mutual exclusivity roles. Cardinality constraint refers to limiting the number of roles that an individual user can belong. Moreover, RBAC supports constraints on the authorization of an operation to a role and context constraints such as location and time on operations being performed on objects. Authorization constraints are an important aspect of access control and are powerful mechanism for laying out higher-level organizational policy. The separation of duties (SoD) is a well-known authorization constraint that reduces the risk of frauds by not allowing a single individual to have complete authority within the system. SoD is implemented in terms of mutual exclusion of roles. RBAC enforces SoD constraints on subject-role relationship and role-permission relationship [4]. In particular, static SoD imposes restrictions on the assignment of subjects to roles. For instance, a subject assigned to the doctor role may not be granted the nurse role. In general, the constraint ($role1, role2, 2$) states that a subject may be granted one out of the two roles as specified in the constraint. Dynamic SoD constraints are enforced at run-time and are outside the scope of this article.

Role management in RBAC can be simply divided into solving the problem of assigning subjects to roles, assigning permissions to roles, and assigning roles to roles to create a role hierarchy, i.e., an inheritance relationship among roles. Moreover, a role hierarchy creates a constraint, e.g., junior child roles obtain the permissions associated with senior parent roles. Consequently, a subject assigned to a parent role automatically obtains the permissions associated with all the child roles. The role engineering process is the first stage of role man-agement in RBAC, which can be used to create role hierarchy from an existing infrastructure. Till now, subject-role assignment, role-permission assignment and role-role relation are all implemented using list style. However, the current implementations of RBAC require two significant challenges: (a) *Constraint Problem*: constraints such as mutually exclusive roles, i.e., a subject can be assigned to at most one role set to achieve the separation of duty constraint. (b) *Redundancy Problem*: Two types of redundancy can occur in the subject-role assignment. First, a subject should not be simultaneously assigned a parent role and child roles because the parent role already contains the permissions associated with the child roles. Second, repetitive permissions should not be assigned to the same user. This can happen when the same user is assigned multiple roles which contains the same permissions. It can be simply solved by revoking that user from one of the redundant roles. Finally, the directed acyclic graph (DAG) used to represent role hierarchy can be visualized to display the structure of RBAC and the assignments. Indeed, such visualization can be exploited to detect mutually exclusiveness and redundancy [5].

Role-based security provides various advantages such as substantially reducing the management cost of access control [5]. RBAC provides lightweight security in the sense that it requires less resource consumption in the blockchain compared to the existing access control systems [6] Moreover, flexibility and expressiveness are the top requirements for an access control system. RBAC is a flexible access control solution because it regulates access to resources based on the subjects' activities and job functions within an organization. Moreover, RBAC supports the reassignment of a subject from one role to another and the assignment of new permissions to roles. Broadly speaking, multiple subjects can use the same role to access resources and a single subject can exercise different roles on diverse occasions. Another advantage of RBAC is that an administrative user controls access to resources at a higher level according to the way organizations perform business. An organization creates roles, role hierarchies and constraints to regulate users' actions. Therefore, once an RBAC framework is established, the administrative user only needs to grant and revoke roles to support changes in the organizational structure. Indeed, this offers ease of administration and a more intuitive process of managing access control compared to the conventional lower level access control mechanism such as access control list (ACL), etc [1]. RBAC can be used to control access to both logical or physical objects[1]. For instance, files, tables and directories are common examples of logical resources that hold critical information. An object can also indicate critical physical system resources such as computational system, printers, disk storage and so on.

Enterprises can adopt the RBAC system to enhance their security requirements. Enterprises can deploy RBAC system to achieve its major benefits as discussed earlier. Consequently, our proposed scalable blockchain-based RBAC is motivated to solve the problem of access control in organizations that do not wish to rely on a third-party service to manage the trust relationship. However, delegation, monitoring of trust relationship and a decentralized design without relying on a third party represent the major challenges to achieve this goal. Unfortunately, existing solutions do not address these challenges well. Access control list (ACL) provides a simple mechanism to meet the security requirements of small organizations containing small number of subjects. However, large organizations demand a flexible, scalable, decentralized and fine-grained access control mechanism because such an environment contains a large number of subjects and objects [7]. ACLs typically associate an object with a list of subjects and groups. A subject performs an operation on an object if that subject or a group to which that subject belongs is specified in the ACL associated with the object. ACL is appropriate for implementing security at a lower level, while RBAC uses a higher level of abstraction (roles), which better serves the needs of organizations. Moreover, RBAC provides more features than ACL such as constraints and role hierarchy. An implementation of the RBAC system requires approximately the same amount of processing as that required by an ACL implementation [8].

---

[1] objects or resources can be contents, services, applications, or computing power. In this paper, we use both terms interchangeably.

A systematic literature review on blockchain in [9] has shown that latency and throughput are amongst the major challenges that are not addressed widely in earlier blockchain platforms. Also, future blockchain solutions will contain millions of people. Transaction latency is an essential characteristic for a blockchain platform. It can be simply defined as the total time taken in executing a transaction and performing its validation. The EOS blockchain requires 0.5 seconds to create new blocks in the network which is comparatively very less than the existing blockchain platforms. Indeed, a large number of people and organizations are now attracted to the EOS blockchain due to its scalability to support millions of people [10]. Transaction fee is also a major concern for blockchain platforms. Existing smart contract platform such as Ethereum [11] is currently facing a major problem in terms of transaction fees. Any transaction in the Ethereum blockchain requires users to pay transaction fee to promote the mining node for invoking the transaction and for adding it to the chain. However, the EOS blockchain has eliminated this problem, thus, enabling users to make gasless transactions without incurring any cost [12]. Parallel performance is also inevitable for a blockchain platform because it processes transactions per seconds. Unlike earlier blockchain platforms [11], the EOS blockchain supports parallel execution of transactions. The EOS blockchain achieves the above performance advantages mainly due to its consensus algorithm. It utilizes the Delegated Proof-of-Stake (DPOS) [13] consensus algorithm which is improved compared to the consensus techniques of the earlier blockchain platforms such as the Proof-of-Work (PoW) [14] consensus algorithm.

We argue that adopting blockchain for access control solutions is not straightforward and involves several significant challenges such as (i) long latency to confirm transactions (ii) low scalability of the existing blockchain platforms such as Bitcoin [15] and Ethereum [11], which is mainly due to broadcasting transactions to the blockchain network and (iii) high resource consumption to solve the PoW cryptographic puzzle. Consequently, we propose to use the EOS blockchain to solve these significant issues. The EOS is the latest blockchain platform which is designed for both private and public use cases. An interesting feature of EOS is that it can be used for business needs across organizations. Moreover, the EOS project was launched to offer an operating system-like environment to enable a quick scaling of decentralized application (i.e., DAPP**2**). Indeed, this platform is intended to overcome the limitations of the existing blockchain platforms. To this end, the EOS architecture has considered various novel techniques to provide higher throughput and processing abilities [16]. It is worth noticing that the EOS blockchain provides a platform that is particularly optimized for smart contracts. The ability of this blockchain to provide transactional throughput in the millions has attracted the attention from both academia and industry.

This paper's contribution is the development of an EOS DAPP to translate Role-based access control policies to smart contract deployed on the EOS blockchain. Our proposed RBAC smart contract accepts different transactions such as role assignment, role revocation, role update, permission assignment, permission update and permission revocation from the issuer as well as policy evaluation transaction from the service/data provider or third parties to vali- date RBAC policies in the blockchain. Our proposed smart contract is deployed and tested on the EOS Kylin tesnet. Our proposed blockchain-based RBAC provides a scalable, decentralized and simple management of roles and permis- sions. Our system provides the advantages of both trust-management and RBAC system to meet the needs of various organizations. Our proposed approach ex- ploits blockchain identities to identify all entities involved in a trust-sensitive operations and to validate delegation certificates. The issuer grants roles to the EOS public keys representing the registered subjects in an organization. We provide qualitative arguments to demonstrate that our proposed blockchain-based RBAC achieves privacy, auditability, availability, integrity and confidentiality and also discuss how our proposed solution thwarts security attacks such as Denial of Service (DOS) attack [17] and impersonation attack. Experimental results and analysis clearly reveal that our EOS blockchain-based RBAC outperforms earlier blockchain platforms such as Bitcoin [15] and Ethereum [11] in terms of cost, latency, throughput, contract execution time and block generation time. Consequently, the EOS blockchain is a promising platform to design scalable and efficient access control systems for several environments. Finally, our blockchain-based RBAC is not limited to organizations since it can equally be used by resource owners to control access to their valuable and critical resources.

The rest of this paper is organized as follows: Section 2 is dedicated to the Background and Related Work. Section 3 discusses the Role-based access control system. Section 4 presents our proposed Blockchain-based RBAC system. Section 5 discusses the advantages of our proposed approach. Section 6 evaluates the performance of the proposed system. Finally, Section 7 concludes the paper and highlights further challenges in this area.

---

[2] (DAPP) is a program that allows users to interact with smart contracts running on the blockchain, and store persistent data on the blockchain.

## 2 Background and Related Work

### 2.1 Access Control System

An access control system (ACS) can be simply defined as a system that protects the critical resources by checking the access requests sent by subjects. An ACS can be designed by considering an access control model. It decides whether the requesting subject has the right to perform the requested operation on a requested object. To this end, the ACS validates the conditions defined in the subject's access control policy. If the policy conditions match with the credentials of the subject, then the subject becomes authorized to perform the requested operation on the requested object. Otherwise, the request is rejected.

An access control system should meet the security requirements of privacy (hiding the personal information of subjects), integrity (preventing the modification of objects), confidentiality (preventing unauthorized access to data) and availability (resources are always available to legitimate subjects) [18]. A complete access control system requires the following main functions:

- Authentication: Authentication identifies a legitimate or registered subject. Different information can be used to accomplish authentication such as the subject's password, private key of the subject, smart card, etc [18].
- Authorization: Authorization is the principal process of an access control system that grants or denies permission to an authenticated subject based on the information in the access control policy. It includes essential steps such as selecting an appropriate access control model, defining the mech- anism to store access control policies and implementing the access control mechanism to enforce the functionalities defined in the access control model. This research primarily focuses on the authorization process by considering and implementing the RBAC system to achieve an auditable and scalable solution exploiting the EOS blockchain platform.
- Accountability: The system should be able to produce logs or traces of the users' actions for auditability purposes. This functionality requires technical solution for log management. Blockchain is a disruptive technology that possesses the potential for tracking users' actions. Hence, making it a perfect candidate for creating an auditable and scalable RBAC system.

Access control system can be classified into four types (i) Discretionary Access Control (DAC) (ii) Mandatory Access Control (MAC) (iii) the famous Role-based access control (RBAC) [19] and (iv) the Attribute-based access control (ABAC) [20]. Discretionary access control restricts access to resources based on the identities of users. Discretionary access control can be characterised from its main feature: leaving access permission as a choice of the resource owner who controls which subject can perform an operation on a certain resource. Operating systems such as Windows and UNIX largely rely on discretionary access control to restrict access to resources. The Access Control List (ACL) provides a simple mechanism to implement these policies in operating systems. That is, DAC associates each resource with a list of subjects who are allowed to access that resource. A subject can perform an operation on a given resource if that subject is present in the access control list. These policies can be flexible, but weak in security. Moreover, DAC cannot control information efficiently because owners can optionally delegate their authorities to other subjects.

Mandatory access control attaches security labels and clearances to subjects and objects to enforce access control decisions. Different security labels are asso- ciated with different privileges. A subject can access a given resource if the sub- ject's clearance is equal to or greater than the object's label. Unlike discretionary access control which allows resource owners to enforce policies, mandatory access control policies are enforced by a higher authority. It creates an access control relationship that the resource owner cannot change. In other words, the admin- istrative user enforces mandatory access control policies for all subjects involved in the environment. Finally, discretionary access control supports tranquillity in the sense that security labels on subjects and objects do not change (except by the administrative user). Military and sensitive applications commonly exploit mandatory access control policies to control access to valuable resources. Manda- tory access control policies are usually focused on preserving the confidentiality of data. However, these policies are expensive and difficult to implement.

On the contrary, Role-based access control exploits multiple roles to sep- arate the responsibilities of subjects. RBAC manages the access control rela- tionship between subjects and objects independently. Consequently, it assigns access rights to roles representing job responsibilities, not directly to subjects [21] [18]. Finally, Attribute-based access control is a logical access control system which considers the attributes of subjects, objects and various environmental at- tributes to control access to resources. ABAC controls access decisions simply by changing attributes values, without the need to change the subject/object relationships. ABAC differs from RBAC in the sense that resource owners or administrative users apply access control policies without prior knowledge of the subjects. ABAC is also flexible because the administrative users do not need to modify objects and rules. When a new subject joins an organization, the admin- istrative user simply assigns the attributes necessary for access control decisions. However, ABAC is not appropriate for sharing information among multiple or- ganizations due to its increased complexity that requires an array of functions and an attribute management infrastructure [20].

## 2.2 Access Control Requirements

Role-based Access Control (RBAC) is one of the most popular and widely de- ployed access control system. In both intra-organizational and inter-organizational workflows, RBAC provides the basis of access control. RBAC provides various well-recognized advantages over other access control systems such as Mandatory Access Control (MAC) and Discretionary Access Control (DAC). Among these, flexibility and ease of management are the greatest virtues of RBAC. The au- thors in [22] have identified a few access control requirements for organizations. This article further investigates the requirements as follows: The access control service should meet the following requirements:

- Least Privilege: Many organizations require that subjects should only acquire the permissions needed to perform their job functions. In other words, this principle requires that a subject should not be given no more privileges than necessary to perform his job functions. To ensure least privilege, a subject's job functions first need to be identified. Then we determine the minimum set of privileges that are needed to perform that function. Non-RBAC systems can solve this problem, but in a costly and difficult way. For instance, consider the case of a capability-based system where a subject assigned to a job function may obtain more privileges than required. This is due to the fact that the capability-based system does not consider attributes or constraints to enforce access control decisions. However, RBAC allows a subject to perform only those operations that are needed once that subject is assigned an appropriate role. Therefore, RBAC assigns role for a particular object based on the least privilege principle [1] [23].
- Delegation of Authority: Employees in an organization should be able to transfer access rights to other employees for some reasons. For instance, a subject designated as a leader of a group of peers in an organization may delegate tasks to peers in that group. To accomplish this goal, RBAC supports the delegation of authority principle [24]. With this property, a subject (del- egator) transfers the access rights to another subject (delegate) to perform an activity on behalf of the former. The delegate should always be a person with reasonable skills and experience. Delegation increases the success of or- ganizations because the whole team becomes responsible to complete tasks. Effective delegation also saves time because tasks are delegated to group members, which decreases delays in accomplishing tasks.
- Reflect the organization structure: The access control service must reflect the structure of the organization. An organization typically expresses this structure in the form of superior and inferior positions such that superior employees have more privileges compared to inferior employees. RBAC supports the role hierarchy feature to meet the structural requirement of organizations.
- Simplified Security Administration: Organizations require that permissions should be assigned to the subjects' job functions rather than to their identities. This is desirable because new employees join the organization and existing employee may leave the organization. Compared to the existing access control systems, the use of roles simplifies the administration of security policies and reduces the administration cost [4] [7]. For instance, if an organization assigns a new responsibility to a subject, the subject's existing role needs to be revoked and a new one will be assigned according to the new duty of that individual. Consequently, RBAC eliminates the burden of creating permission for each employee in the organization.
- Separation of Duty Principles: All organizations should separate functional responsibilities. Organizations must enforce the separation of duty principles to deter the possibility of fraud. This is due to the fact that proper separation of duties is vital for the implementation of effective access control. RBAC supports the separation of duty principles which assures that mistakes, intentional or unintentional, cannot be made without being discovered by another person [4]. These principles may be enforced at the design time or during run-time. Design-time enforcement prevents circumstances that should never occur. For instance, a financial manager should never be allowed to be an auditor as well. However, run-time enforcement ensures that permissions are not misused.

- Policy-Neutrality: Many organizations require policy-neutral access control systems that can express a wide range of security policies. RBAC is widely regarded as a policy-neutral system because it has the potential to express a wide range of access control policies by using permissions, constraints and role-hierarchies [4]. The feature of policy neutrality means that RBAC can be configured to enforce different policies, such as discretionary access control (DAC) and mandatory access control (MAC). Osborn et al. [25] have presented a systematic construction to generate both of the traditional access control mechanisms from the standard RBAC. It is worth noticing that RBAC components such as subjects and permissions are necessary to express any access control system. RBAC can be used to generate DAC policies because DAC also rely on a large number of administrative roles, which play an important role in enforcing DAC policies. Similarly, the role hierarchy feature plays a vital role in generating mandatory access control e.g, the lattice-based access control (LBAC). Thanks to the policy-neutrality feature, RBAC is not confined to a specific single organization.

- Special Administrative Roles: Each organization is typically partitioned into several departments such that each department manages its own functions independently. RBAC can be configured to define special administrative roles to meet the requirements of such organizations [26]. An administrative role is a special type of role that describes the ability of the subject belonging to that administrative role to manage role assignments. This characteristic of RBAC makes it a powerful system to meet the needs of diverse scenar- ios including the Internet-of-Things (IoT), Social Networks, Healthcare and so on. Such large systems consist of multiple domains or units such that a special administrative role will be responsible to manage each domain independently. These special administrative roles will independently perform the management of roles and permissions.

- Improved productivity of organizations: Many organizations demand an efficient access control system that requires reduced cost. RBAC has the potential to enhance organizational productivity by improving the systems that organizations use to structure their information systems. Thanks to the flexibility and the higher level of abstraction, the RBAC system can efficiently support changes in the organizational structure. RBAC promotes new ways of structuring the organization, thereby affecting the organization's productivity. Finally, RBAC supports the management of access control policies across different divisions or units within the organization such that each division or unit has its own administrative user who can assign/revoke roles/permissions [7]. Indeed, several administrative roles can be defined, depending on the organization's requirement, to efficiently manage the policy management task.

## 2.3 Blockchain

The Bitcoin was the first cryptocurrency which exploited and emerged the blockchain to the world in 2009 [27]. Bitcoin is essentially a decentralized cash system to transfer money in a peer-to-peer (P2P) mode without relying on a third party (i.e., a bank). Blockchain technology was initially developed to pro- vide cryptocurrency services in a secure way without depending on a third party. Thereafter, many alternative cryptocurrencies (electronic cash) have entered the scene using similar structures. Nowadays, this technology is exploited for applications beyond cryptocurrencies.

*Blockchain* is basically a distributed ledger technology that accepts transactions from the users and organizes these transactions into blocks. Each block contains a set of transactions. These blocks are broadcasted to the P2P blockchain network where they are confirmed by the participating nodes in the blockchain network. The newly approved block is appended to the existing block in such a way that the new one contains a cryptographic hash of the directly previous block. Indeed, the hashing process is essential to improve the security of the blockchain because any alteration in a single block will change the hashes for the linked blocks as well. Consequently, conflicting copies of the shared ledger will emerge and the state of the blockchain will become inconsistent. The term shared ledger is commonly used to describe the blockchain because the ledger is shared among the peers in the blockchain network such that each peer is capable to directly access a copy of the distributed ledger. It is worth noticing that blockchain uses a synchronization process through which a node communicates with other nodes to request and import the latest blockchain data. The pos- sible applications of blockchain technology go beyond Bitcoin. The blockchain technology has the following characteristics.

- Data Auditability and Transparency: Every executed transaction and the data structures stored in the blockchain are publicly visible to all peers in the network.
- Decentralized: A decentralized solution which is operated without the support of a central authority.
- Consensus: Instead of a central authority, the participating nodes in the blockchain network validate all the broadcasted transactions.
- Information Distribution: The information is distributed to each node in the network to ensure availability and consistency.
- Security: Malicious users cannot misuse the blockchain because it is tamper-resistant.

Security, auditability and a decentralized design are the principal qualities of blockchain technology. These features make the blockchain and ideal candidate to design a scalable Role-based access control solution for organizations as well as for the resource owners.

A *smart contract* is a program to perform a specific task [28]. Basically, a smart contract contains a program code and a storage file. A contract may read or write data to its storage file when users send transaction to it. The user writes the programs, compiles it and then deploys it on a smart contract platform. Besides Ethereum, a number of alternative blockchain platforms have emerged over the last few years that support the functionality of smart contracts [29]. For instance, Stellar, Monax, and EOS are the new emerging platforms that enable users to write smart contracts to accomplish different tasks. It is worth noticing that each platform uses different programming language to write the program code. A contract creation transaction is usually used to deploy a new contract on the blockchain, and the contract logic is executed by sending transaction to it. Moreover, such transactions are invoked by the participating nodes in the blockchain who reach agreement on its output and modify the blockchain accordingly.

*Consensus* is a challenging problem in distributed systems that involves a set of agents to jointly decide a given value required for computational tasks. Each blockchain platform exploits a consensus protocol to reach an agreement about the next block to be included into the blockchain. It is worth noticing that a trustworthy consensus process needs reliable agents in the network. Different consensus algorithms have been proposed for blockchains platforms including the Proof-of-Work (PoW), Proof-of-Stake (PoS), Delegated Proof-of-Stake (DPoS), Stellar Consensus Protocol (SCP) and Ripple Protocol Consensus Algorithm (RPCA). Each consensus mechanism has its own advantages and limitations in terms of energy consumption, security and power consumption. The DPoS algorithm is energy-efficient and secure compared to the PoW algorithm [14]. The PoW decentralized consensus algorithm requires a certain level of difficulty to produce new blocks in the blockchain network. This mechanism checks the validity of each broadcasted block in the blockchain network. The miners reach a common agreement on a newly broadcasted block and add it to the chain. The added block cannot be modified due to the immutability property of blockchain. On the contrary, the latest DPoS algorithm [13] exploits participants' votes to elect the block producer (BP) node to generate new blocks. The system uses maintenance interval to shuffle the BP. Each block producer can generates 6 blocks with a block time of 0.5 seconds. If a block producer misses a block, then simply an empty block is appended to the chain. It is worth noticing that the system removes a BP from consideration if it has not generated any block within the last 24 hours.

Furthermore, cost, consistency, extensibility, security, performance and scalability are the critical success factors affecting the decision to adopt the right blockchain technology for solving problems [30]. A blockchain network imposes certain costs for running the network. For instance, transaction fee indicates a variable cost to run transactions, smart contract deployment cost and incen- tives cost for processing transactions. We note that transaction fees in Bitcoin have raised up to 40 USD. Ethereum uses a cost for instruction approach and it then calculates the overall cost of transactions. However, the EOS implementa- tion does not charge transaction fees while performing millions of transactions per second [12]. *Consistency* is a function of the time to confirmation (e.g., 60 min for Bitcoin, 1 minute for Ethereum and 1 sec for EOS). Different tech-niques have been used to ensure that a transaction is added to the blockchain. A transaction is appended to the blockchain after generating a certain num- ber of blocks. For instance, the Ethereum platform waits for the generation of 12 blocks after adding a transaction into the chain. Consistency primarily de- pends on the block production rate (BPR). Possible values for time to confirma- tion are: seconds, minutes and hours. Possible values for block production rate are: 10 minutes or more, 1 to 10 minutes and seconds. *Extensibility* determines whether the platform is extensible through smart contract and decentralized ap- plications (DAPPs). Bitcoin has a limited support for decentralized applications and smart contracts. However, Ethereum and EOS platforms are extensible be- cause they support the functionality of these additional useful features. Some blockchain platforms provide ad-hoc languages such as *Solidity* in Ethereum to support turing-complete smart contract. On the contrary, other platforms such as the EOS supports traditional languages to write the code of a smart contract. *Scalability* refers to the capability of the blockchain to server more users and transactions. For instance, the PoW algorithm offers good scalability with poor performance, whereas the DPOS algorithm offers better scalability with better performance. Finally, blockchain provides security in the sense that the data on the public ledger cannot be deleted or modified once it is written to the chain. An attacker may exploit the vulnerabilities in a blockchain platform to launch an attack. The DPoS algorithm provides resistant to several attacks including short range attack and selfish mining attack. However, the PoW algorithm is vul- nerable to selfish mining attack, Sybil attack [31] and Denial of service attack [32]. It is worth noticing that the most popular Bitcoin blockchain platform is designed without the functionality of smart contracts. However, alternative im- plementations that support this functionality are currently running and some are in development. Moreover, blockchain implementations vary in many ways such as their purpose, cost and efficiency, among others. The choice to decide which blockchain platform should be adopted for a given application, it is necessary to recognize the differences among them. Table 1 shows a comparative analysis of the popular blockchain platforms in terms of interchain communication, allowing multiple parallel blockchains to interoperate retaining their security properties and other essential features.

Table 1: Comparative analysis of the popular blockchain platforms

| Features | Bitcoin | Ethereum | EOS | Stellar |
|---|---|---|---|---|
| Smart Contract | Limited | Yes | Yes | Limited |
| DAPP | No | Yes | Yes | No |
| Languages | C++ | Solidity | Any | Any |
| Confirmation Time | 60 min | 1 min | 1 sec | 30 seconds |
| Block Production Rate | 10 min | 10-19 sec | 0.5 sec | 5 sec |
| Consensus Algorithm | PoW | PoW | Delegated PoS | Stellar Consensus |
| Interchain | No | No | Yes | No |

## 2.4 EOS Blockchain

EOS is a collection of applications that interact with the blockchain. The EOS blockchain platform for developing smart contract and DAPP aims to address the scalability problem found in existing blockchain platforms such as Ethereum and Bitcoin. This smart contract platform aims to eliminate transaction fees while conducting millions of transactions per second. The EOS blockchain is de- signed as a peer-to-peer blockchain network that utilizes optimized techniques to function in a particular execution environment. The consensus mechanism in this blockchain largely depends on the 21 block producers. These block pro- ducers are mainly responsible for computation on the data and for coordinating updates across the blockchain network [16]. Moreover, they process transactions in the blockchain network to reach a common consensus among all the indi- vidual databases within the EOS network. This platform relies on the C++ programming language to write the code of smart contracts and it possesses the high potential for scaling due to the Delegated Proof-of-Stake (DPoS) consensus mechanism [33].

The core components of EOS blockchain include *Nodeos*, *Cleos* and *Keosd*. *Nodeos* validates/syncs blocks in the network. EOS classifies nodes into producing nodes (block producers) and non-producing nodes that are associated with EOS accounts [16]. Developers can exploit the remote procedure call (RPC) API to implement non-producing nodes. Moreover, these nodes can also serve as private endpoints for applications that require the EOS system for local development. *Cleos* provides a command line interface allowing clients to interact with the EOS blockchain by sending transactions. Clients request resources from the block producers (BPs), which are primarily responsible to host, deliver and manage resources on a demand basis. The *Keosd* component runs on local computers to stores private keys locally. Keosd is essentially a key manager service daemon for storing private keys and signing digital messages. It provides a secure key storage medium for keys to be encrypted in the associated wallet file [33]. Figure 1 shows the basic architecture of EOS blockchain. We now discuss the essential elements of the EOS blockchain in the following section.

EOS account names consist of short convenient usernames that are at most 12 characters in length [34]. An EOS account is basically a set of public and private key. The public key serves as the home address for a user to receive tokens or transactions and it can be openly and freely shared with the community. However, the private key is secret and only known to the owner of the EOS account. Consequently, an EOS account is controlled by the private key, has EOS tokens, and used to perform all the transactions in the EOS blockchain network. These transactions, in turn, perform some actions on the blockchain network by executing methods of the smart contract. It is worth noticing that key generation is essential to create accounts, where the generated keys are associated with their corresponding wallets. The Cleos application provides instructions to perform this operation. Moreover, all the components operate together to publish the newly created account within the network. Cleos sends request to Nodeos to generate accounts and publish them into the network. Users import their private keys into their wallets using the Cleos tool. Consequently, EOS exploits the keys associated with users' accounts to sign all the actions performed in the network. Moreover, when creating multiple wallets, a user must import the required keys into these wallets to successfully create signatures for signing transactions. Finally, EOS accounts exploit trust techniques to validate actions. Block Producers will be able to suspend an account if it violates this trust mechanism. Indeed, trust mechanism is the only option which is used by the consensus protocol to eliminate bad actors from the network [16].

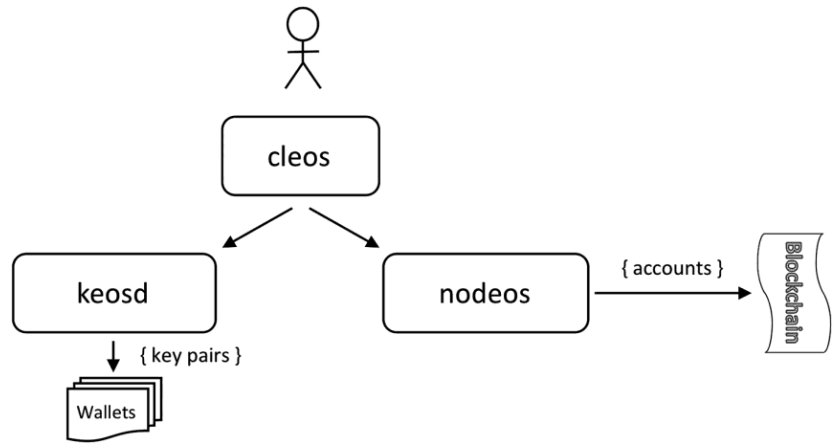

Fig. 1: Basic Architecture of EOS blockchain

Within EOS blockchain system, a transaction consists of a set of instructions to modify the state of the chain. An EOS transaction essentially represents the execution of actions. Transactions can be sent to perform a single action or multiple actions. The state of EOS blockchain will change when a transaction is successfully executed. A transaction sent to execute the function of a smart contract performs some action according to the instructions in that function. However, to execute the function of a smart contract, users must be registered and associated with EOS accounts. A transaction may specify some input parameters when executing the function of a smart contract.

The EOS blockchain has created a more robust architecture than Ethereum for smart contract and DAPP development. A smart contract in the EOS environment consists of variables and methods that perform operations on the variables. The methods of an EOS smart contract can be executed by sending action requests that are stored within the EOS database framework [33]. The EOS platform compiles the source code of a smart contract into bytecode so that the EOS virtual machine can execute it. Moreover, the EOS blockchain defines smart contracts in terms of action and action handlers. An action is es- sentially a call to a method of smart contract whereas action handlers performs the tasks of the requested actions. Actions can also be sent to access the private database associated with a smart contract. For instance, our RBAC smart con-tract has an associated database containing role-based access control policies. A storage device, service provider or even third parties send actions to check access permissions in the blockchain.

Finally, the EOS blockchain supports the C++ programming language to write the code of smart contracts. This code will be compiled down to Web Assembly (i.e., WASM) format and then invoked in the EOS virtual machine [34]. WASM can be used to compile the code for various high-level languages such as Rust, C and C++. In simple words, WASM provides instructions that the EOS virtual machine recognizes. Moreover, WASM is a web standard specifying the binary instruction format for a stack-based virtual machine [16] [35]. Thus, it can be executed in several environments including the latest web browsers.

## 2.5 Related Work

To the best of our knowledge, a few Blockchain-based access control systems have been recently presented, which exploit either the Ethereum or the Bitcoin platform for access control decisions. Maesa et al. [36] developed a parser to translate Attribute-based access control policies to smart contract. They use smart contract to facilitate the process of policy creation, revocation and evaluation using the public Ethereum blockchain network. However, the Ethereum blockchain is not scalable and transactions are not cost-free.

Third parties should not be trusted for storing users' personal sensitive data because they are vulnerable to misuse and attacks. Zyskind et al. [37] merge the blockchain and off-blockchain storage to design a personal data management platform which is particularly privacy-centered. This platform primarily protects the privacy of mobile users who download various third-party applications for different purposes. These mobile applications constantly collect high amount of personal data from the users without their knowledge and control. Their proposed platform protects against different types of privacy problems such as ensuring data ownership privacy to allow users to own and control their own personal data, ensuring data transparency to make the data collection transparent to the users, and ensuring access control to control access to the users' personal data. Their proposed solution is simply based on two types of transactions: access transactions control access to resources whereas data transactions are used for data storage and retrieval. It is worth noticing that they use a pointer to the data on the blockchain and an off-blockchain key-value store (a distributed hashtable) to process data transactions. However, the authors simply propose a theoretical access control solution without concrete implementation. In other words, their work is mainly focused on the read operations of the users' stored data, without focusing on the access control problem for users' sensitive data. Moreover, their solution is based on the Bitcoin blockchain platform, which has a major scalability problem.

A. Ouaddah et al. [38] use the computing capability of blockchain to enforce access control decisions. Their scheme uses the blockchain as a decentralized access control manager. Access tokens, which represent authorization, can be transferred from one user to another using transactions. Moreover, they insert access control policies into the locking scripts of transactions. The receiving user needs to unlock the locking script to verify the ownership of the token. However, tokens are not implemented using smart contracts and locking scripts have a limited computing capability. It is worth noticing that their proposed access control system is based on the organization-based access control (OrBAC). However, OrBAC has several drawbacks such as lack of support to meet the interoperability, collaboration and distribution needs for distributed structures, which are critical aspects in distributed environments. Therefore, it can handle only policy-based compatible systems and use cases, which cannot be applied to numerous IoT contexts. Though their solution takes simple context information such as time and location into consideration, it does not support constraints such as separation of duty and advanced context attributes such as date, birthday, affiliation to an organization, employee's salary or nationality. Finally, their solution is based on the Bitcoin blockchain platform which has a major scalability problem.

Ali et al. [39] propose an access control solution exploiting blockchain technology. The main actors of the proposed smart home scenario include a cloud storage, service provider and user devices. Access control policies are stored in the policy header of a local private blockchain. Each home miner manages the outgoing and incoming access requests to the smart home. Furthermore, smart homes form an overlay network that contains users' devices, cloud storage together with the service provider. The overlay network makes their proposed architecture distributed. All the nodes in the overlay are groups into clusters such that each cluster maintains a cluster head, which is responsible to main- tain a public blockchain. Each smart home device relies on the cloud storage to share data in the network. However, the overlay network further increases network delay and overhead. Moreover, the private blockchain operates with- out the Proof-of-Work (PoW) consensus algorithm, which is necessary to ensure the trustworthiness of the access control technique. Also, the authors did not provide detailed experimental results such as throughput, latency and resource consumption. It is worth noticing that their solution rely on simple policies without taking constraints and exceptions into consideration. To eliminate the overlay overhead and introduce trustworthiness using the DPoS protocol, each home owner will create RBAC policies using his personal RBAC smart contract deployed on the EOS blockchain. In addition, each smart home would execute the check access function of our proposed RBAC smart contract to check permissions for the incoming and outgoing access requests. For better availability, data can be stored in the distributed Interplanetary File System (IPFS). Our RBAC smart contract first checks the subject-role relationship to fetch the role assigned to a requesting subject. It then references the role-permission relationship to identify the permission (operation, object). In this case, operations can be either store, monitor, access, etc., while object indicates the ID of the target device. The check access function of our RBAC contract returns a BOOLEAN true or false decision, representing allow or deny, respectively.

Zhang et al. [40] propose a smart contract-based framework consisting of multiple access control contracts, one judge contract, and one register contract, to achieve a trustworthy access control systems. However, the access control contracts are based on a simple Discretionary Access Control (DAC) mechanism which incurs high administrative overhead. Moreover, the Ethereum platform has a major scalability problem. The architecture of their proposed framework consists of a server, storage device, user devices, IoT gateway and IoT devices. The server provides various services for users by interacting with the storage device and IoT devices. The storage device stores data for sensors, users and server. A user can use different devices such as laptop, PC and smart phone to access services from the server and read or write data to the storage device. Each IoT gateway uses a short-range communication technology such as Wi-Fi, Bluetooth, etc. to connect a group of IoT devices. Finally, IoT devices primarily include sensors, which observe environmental data and send it to the storage device or server for further processing. It is worth noticing that this scheme incurs high storage and communication cost because the access control list does not consider subjects' information in the policies. Consequently, they have stored an additional register contract in the blockchain, which requires the ID of the resource owner and returns information about the access control contract deployed on the blockchain.

Rahman et al. [41], exploit the Ethereum Blockchain to implement a cost-effective, flexible, auditable and trustworthy role-based access control system to control access to resources. The system was implemented as a smart contract and deployed on the Ethereum Rinkeby testnet. The resource owner uses the public key of the subject to create auditable RBAC policies, while the corresponding private key is used to decrypt the private data of the resource owner. The sys- tem is auditable because the policies are transparently stored in the blockchain. However, the Ethereum blockchain is not scalable.

The increasing convenience of mobile healthcare has raised several challenges for healthcare providers and policy makers [42]. Mobile Ad hoc Network (MANETs) and Body Sensor Network (BSN) provide the foundation to integrate mobile healthcare with our daily lives. These network supports the mobility of patients. However, the mobility of patients raises new challenges for access control in mobile healthcare [43]. Received Signal Strength Indicator (RSSI) can be used to obtain patients' positioning information. Moreover, we need to investigate which mobility model will provide better accuracy as patients move from one location to another location [44] [45] [46].

## 3 Role-based Access Control (RBAC)

RBAC simplifies the complicated form of an organization's access control policy. RBAC exploits the various roles in an organization to perform the access con- trol decisions. RBAC is a form of non-discretionary access control in which the issuer grants a role's permission to a subject by assigning that subject to role. The issuer gives access permissions to roles, and subjects are endowed with roles according to their job functions and tasks/obligations. RBAC represents access rights as *authorization policies* which describes what actions a subject is permit-ted (or forbidden) to perform on a set of target objects. Moreover, tasks/duties are represented as *obligation policies* which describe what actions a subject must or must not perform on a set of target objects. Roles can be assigned to a subject and revoked to support change in the organization structure, without changing the policies [47]. RBAC efficiently manages access control policy because permissions are associated with roles, not with subjects. Many variations of RBAC are proposed in the literature, but the basic architecture of RBAC is that roles are granted to subjects and permissions are associated with roles (not directly with subjects) [1].

### 3.1 Role-based Access Control Components

The essential components of RBAC include subjects, roles, objects and permissions [48]. Individual users in RBAC are called subjects. A subject can be a human being, a mobile device, a peer, a robot, etc. A role or title is often used to represent the capability of a subject to access resources. An object can be simply defined as an entity that receives or holds some information. Permission indicates the specified mode of access to a resource in the system. RBAC is based on the concept that permissions are linked with roles and subjects are made members of suitable roles present in an organization so that they can access the objects in that organization [19].

The basic RBAC comprises subject-role relationship and role-permission relationship to define the access structure. An effective implementation of RBAC requires a thorough understanding of both pairs of the RBAC 3-tuple, i.e. subject-role and role-permission relationships. These many-to-many relationships represent the access control policy of RBAC. The issuer decides these types of relationships. Moreover, these essential components accumulatively determine the access rights of a particular subject to access a particular resource. When a subject sends an access request, the subject-role relationship is referenced to determine which role is assigned. Moreover, the role-permission relationship is referenced to decide the permission associated with an assigned role. Subject-role relationship represents collection of subjects and roles. Subject-role relationship operation grants role to a subject and maintains the subject-role relationship database that contains information about the roles assigned to subjects [49]. Subject-role relationship is a critical administrative function which is decentralized and delegated to subjects. The issuer directly configures these components or defines administrative roles to configure them. Subject-role relationship can be easily modified to meet the changing needs of organizations. Sandhu et al. [50] have proposed an RBAC system called URA97 that extensively focuses on subject-role relationship. They claim that URA97 can be applied to any RBAC system because subject-role relationship is a fundamental administrative function which will be necessary in any RBAC system. However, URA97 is designed to have a very narrow scope. It is worth noticing that subject-role relationship needs to be decentralized to make the system scale to large systems. To provide greater flexibility and granularity, modern RBAC supports two relations: subject-role relationship and role-permission assignment [21]. To explain the role-permission relationship in detail, we first provide an overview of the structure of permission in modern RBAC.

Permission (an *operation, object* pair) consists of obligation and authorization [51]. According to [47] [21], obligation and authorization expressions are represented as follows:

**identifter, mode, role, action, target, constraints, exception**

[identifier]: uniquely identifies the permission
[mode]: a:authorization, o:obligation, +:positive,-:negative
[role]: role which can exploit this permission
[action]: operation that the role can perform
[target]: represents the target object to apply the operation
[Constraints]: restrict permissions based on time period or working day
[exceptions]: other conditions (exceptional)

The role-permission (R-P) relationship applies permission to a role group. The permission is uniquely *identified* by an Identifier. The *mode* of the policy represents authorization such as positive authorization (permitted: A+), neg- ative authorization (forbidden: A-) and obligations such as positive obligation (must: O+) and negative obligation (must not: O-). In general, a positive pol-    icy is represented by +, while a negative is represented by -. *Role* indicates    the type of role that can use this permission. *Action* represents the operation   that the role can perform. *Target* represents the object influenced by action in permission. *Constraints* are the essential feature of RBAC because they restrict permissions depending on the information provided, e.g., a specific weekday or    a time period. *Exception* represents exceptional condition for permissions to be executed or not, e.g., if R-P contains a negative (－) mode then the specified   role cannot execute this permission, but if some specified exceptional condition occurs then permission will be executed. Exceptional conditions could be, for example, an emergency or accident role. For instance, Bob can specify an emer- gency exception with negative mode so that subjects working in the emergency department could perform the operation listed in Bob's role-permission rela- tionship. On the contrary, if R-P contains a positive (+) mode then only the specified role can execute this permission. Finally, obligation such as "a doctor must read the patient's document (e.g., file3) every afternoon at 1:00 PM" can be simply expressed as *dc1, o+, doctor, read, file3, every 1:00, -*. In addition, an authorization such as "a student can read an article (e.g, file7) of a professor" can be represented as *st1, a+, student, read, file7, -, -*.

A study [52] conducted by NIST demonstrates that the roles assigned to subjects change frequently due to job promotion, departure and arrival of new employees. When a user's  job responsibility changes within an organization,    a proper mechanism is necessary to simply delete membership in the existing  role and grant a new membership according to the new responsibility of that   user. Similarly, when a person leaves an organization, all roles that are assigned  to that person need to be revoked. However, role-permission relationships are usually predefined and change less frequently. Consequently, both assignment operations should be made as simple as possible so that these require less tech-nical skills. Indeed, smart contract can be programmed to define user-friendly functions to accomplish these tasks. The term user-friendly means that the issuer only needs to specify, for example in a *Role Assignment Transaction*, the ID of the subject and the role that needs to be granted. The NIST study team met with 28 organizations to know about their security requirements. They argue that the role-based access control is a better choice for organizations which experience a large turnover of personnel. They also argue that privacy problems were considered particularly critical in several organizations such as healthcare, educational and financial and banking institutions.

## 4 Blockchain-based RBAC System

Our approach follows the role engineering process presented in [22] to identify roles in an organization. First, the organization is partitioned into different units or departments according to the responsibilities of these departments. Second, each department is then partitioned into different divisions according to the types of jobs in that department. For instance, the finance department could be partitioned into divisions such as accountants, secretaries and managers. In the third step, the different positions are created as roles. In the fourth step, superior private roles are identified where the permissions are not inherited upwards. Private roles are superior to their corresponding positions in the role hierarchy. Fifth, task-based roles need to be identified which represent the tasks or duties that individual users must perform. In the next step, subjects are assigned suitable roles according to their positions and duties. Finally, permissions need to be assigned to roles to allow subjects in the organization to perform operations on resources.

Enterprise and subjects' policies represent the two main types of policies in various enterprises. An enterprise policy describes the enterprise subject-role relationship, role-permission relationship and role-hierarchy. Individual subjects define their own RBAC policies containing the subjects' role-permission relationship and constraints. Figure 2 shows the architecture of our proposed scalable RBAC system using the EOS blockchain technology. Each enterprise has a unique RBAC contract deployed on the EOS blockchain that accepts policy ownership transactions only from the issuer in that enterprise. Our RBAC contract contains the necessary functions to allow the issuer to create, update and revoke these policies in the blockchain. Most importantly, each resource owner in an organization may have some resources that are needed by other subjects (in the same or different organization). Consequently, all resource owners must control access to resources to prevent unauthorized access to resources. The service provider should be able to restrict the access requests from unauthorized subjects to perform operations on the stored data. An organization requires scalable role-based access control policies because many subjects are involved in it. Obviously, the Ethereum blockchain is not suitable due to scalability issue and transaction fees. Technically, conventional access control mechanisms can only be used to meet the security requirements of enterprises containing small number of subjects. As a result, conventional mechanisms are not sufficient and complex to manage when we consider a large system involving many subjects from different organizations. To support autonomous policies for subjects, registered subjects associated with EOS accounts send transactions to the RBAC smart contract to delegate access rights.

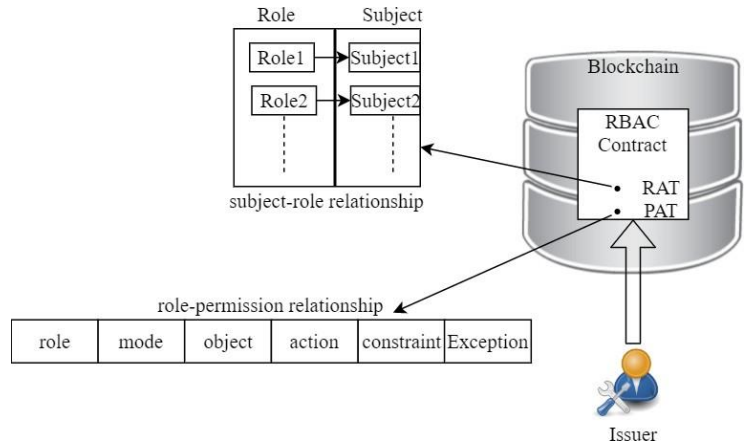

Fig. 2: Scalable Managment of RBAC policies using the EOS blockchain

Figure 3 shows the blockchain-pull architecture of our proposed RBAC system. It consists of an Issuer, subjects, service provider and the EOS blockchain system. The issuer in an organization creates RBAC policies for all subjects working in that organization. The issuer sends transactions to the blockchain to create, update and revoke these policies depending on the requirements of the organization. The subject-role relationship defines which subject is assigned to which roles in which organization. The issuer sends a transaction to grant a role to a subject when the former agrees to create an RBAC policy for a sub- ject working in the organization. A subject sends access request with his/her ID
(EOS public key) to the service provider to perform an operation on a certain resource. The service provider first authenticates [3] the subject and it then pulls subject's role authorization information from the blockchain as shown in Figure 3.

Although we propose to store objects in the service provider, our approach is also feasible if the resources are stored in the Interplanetary File System (IPFS) [53], which is a distributed peer-to-peer (P2P) file system that shares the same system of files with all computing devices. Moreover, IPFS provides a high-throughput storage medium with content-addressed hyperlinks. IPFS uses a combination of technologies such as distributed hash table (DHT) and self-certifying namespaces, etc. The main advantage of IPFS over the service provider would be no single point of failure, and without requiring nodes to trust each other. Another advantage of IPFS over service provider is that the data is stored in a distributed way in various places of the world. Storing and retrieving files from an IPFS work in the same way as the Web. Each uploaded file has a unique hash string which can be used to retrieve that file. The hash string essentially resembles the uniform resource locator (URL) used in the web today. The hash string can be seen as the location of the uploaded file. It is worth noticing that blockchains are not appropriate for storing large files due to gas fees and transparency property of current blockchain platforms. In case when files are stored in the IPFS, our blockchain-based RBAC could require a few steps to check access permissions in the blockchain. When a subject sends a request to perform an operation on a certain resource, the IPFS will use the traditional RBAC triplets to check access permissions in the blockchain. Therefore, the IPFS will return location of the file only if the subject has been successfully authenticated and authorized. It is important to note that IPFS could be an ideal choice to store data if several organizations collaborate to accomplish a common goal.

---
[3] The authentication mechanism is not described in this paper, since it can be supported by any existing authentication technologies in different ways

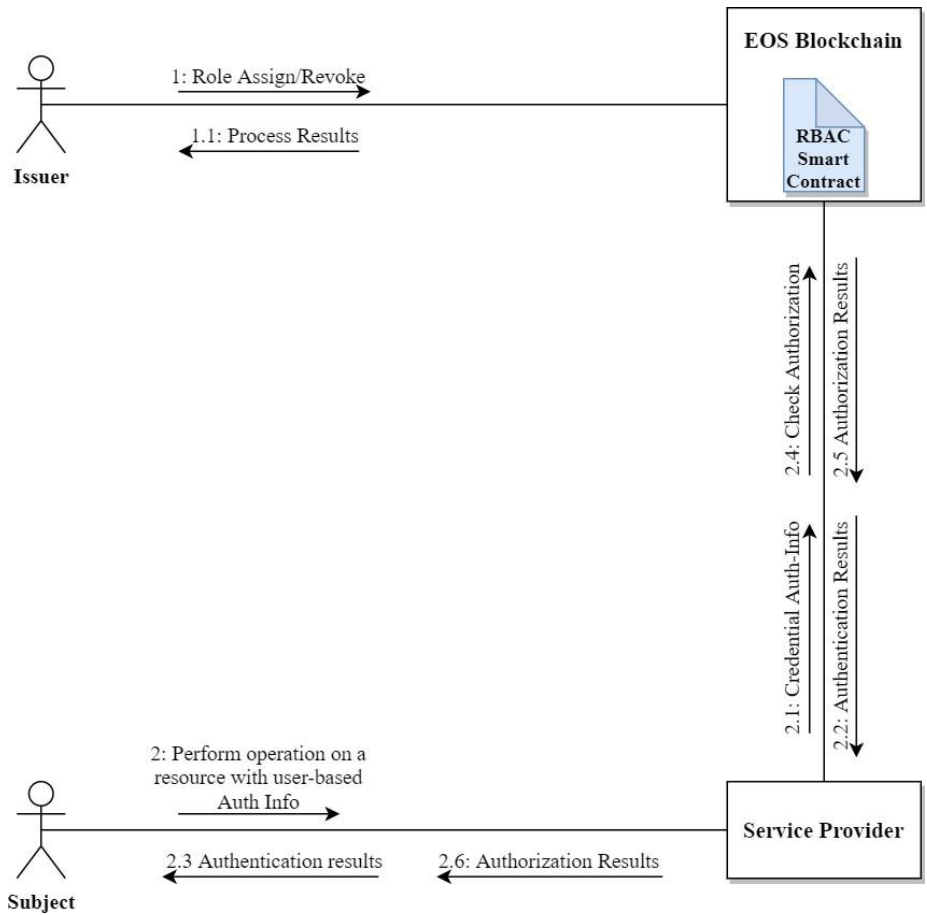

Fig. 3: Block Diagram for blockchain-pull architecture

Finally, the **Check Access** function implements the access control function of our blockchain-based RBAC system. This function requires the traditional access control triplet (subject ID, operation, object) as input attributes. This function first checks the subject-role relationship in the blockchain to verify the role membership of the requesting subject. Moreover, it checks the associated role-permission relationship to determine access permission in the blockchain. This function returns a BOOLEAN true or false, representing allow or deny, respectively. Finally, our blockchain-based RBAC provides a clear interface to other components that use RBAC as their access control service. Figure 4 shows a generic access request scenario as message sequence chart. A subject (left side) sends a request to perform an operation on a certain object (right side).

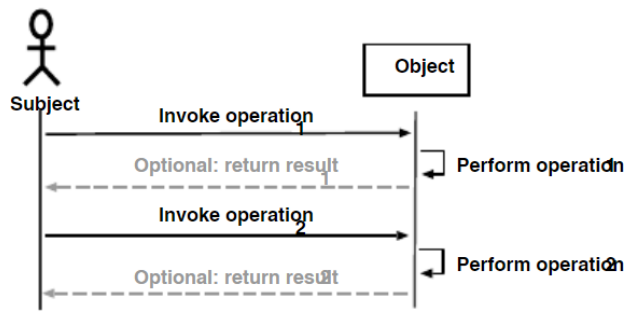

Fig. 4: A generic access request scenario

### 4.1 Delegations

Some environments require that users should be able to transfer the access rights to other users for different reasons. The basic concept behind the delegation mechanism is that some active entity in a system assigns the access rights to another entity to perform some functions on behalf of the former. Delegation has been continuously recognized as an essential concept to increase flexibility in controlling access to resources. Our proposed blockchain-based RBAC supports delegation capabilities using gaseless transactions which makes it adoptable and appealing in a large number of application scenarios. Our RBAC smart con- tract supports two types of delegations (i) administrative delegation and (ii) user delegation [24]. An administrative delegation requires the services of an administrative user (issuer) to only grant access rights to a subject. The term only grant means that the administrative user cannot use the access rights, but only assigns it to a subject. On the contrary, a user delegation enables a subject to transfer a subset of his available access rights to another subject. For instance, a subject may delegate the access rights to another subject when the former is not able to perform a given task due to several reasons such as sickness. Unlike administrative delegation, a third-party delegation operation requires that the subject performing the delegation must be capable to use the access rights. The subject who performs a delegation is referred to as a 'delegator' and the subject who receives a delegation is referred to as a 'delegate'.

Our RBAC smart contract currently supports grant delegation operation because the delegated access rights are available to both the delegator and delegate. On the contrary, transfer delegation operation transfers the access rights to the delegate such that the delegator is no longer able to use the access rights. Transfer delegation operation is necessary in scenarios where the access control policy specifies that the right r needed to perform a task should be available to a certain number of users. For instance, consider an access control policy that demands the cooperation of $n$ users to perform a given task. However, if at least one of the users is not available due to several reasons to complete the task, then users should be able to delegate the access rights to another user, who may perform the task on behalf of the unavailable user. Indeed, grant delegation cannot be used in such scenarios because it will violate the rules of the policy. Consequently, transfer delegation becomes necessary to accomplish delegation in such scenarios [54].

Our blockchain-based RBAC supports temporary delegations such that subject $S1$ delegates to subject $S2$ the role $R$ with an expiry time. Consequently, a temporary role delegation operation *delegate(S1, S2, R, expiry time)* guarantees that the delegation will only remain in the blockchain for the time specified in the expiry time field. However, our simple implementation can be modified to *delegate(S1, S2, R, start time, end time)* or *delegate(S1, S2, now(), now() + T)* where now is a function that would return the value of the current time and $T$ represents the duration of delegation in the blockchain. It is worth noticing that we take a *time* variable in the smart contract to compare the current time with the expiry time field. Our RBAC contract automatically deletes the delegation from the blockchain when the current *time = expiry time*. Nonetheless, a subject can also perform permanent delegation operation without specifying the expiry time field.

If a subject needs the delegation of a role, the delegator sends delegation request message to the blockchain specifying the address of the corresponding RBAC smart contract. The delegation request message contains the role that needs to be delegated, expiry time, and delegate ID which obtains the new role due to the role delegation mechanism. When the blockchain receives the delegation request message, it references delegation information to decide whether to allow the delegation or not. The delegation information consists of permission which has mode and exception. Our blockchain-based RBAC smart contract references the mode for making a decision about whether to permit the delegation or not. The mode of a permission is one of A+, A-, O+ or O-. Permissions with obligation mode O+ and O- cannot be delegated. In other words, right permission can be delegated to other roles, but a permission containing a duty cannot be delegated [21]. Our blockchain-based RBAC smart contract will either accept or reject the request based on the mode of the permission. If the request is accepted, the blockchain sends an accept message to the delegator containing information about the delegator, delegate, role, expiry time and condition. The blockchain smart contract then sends notification of the delegated role to the delegate. The delegate can perform the function of the delegated role once he obtains delegated role information. At this time, the delegate can execute the permissions that are specified in the corresponding role-permission relationship which is stored in an auditable manner in the blockchain.

After a delegation role has been successfully assigned to a subject, it is stored in the blockchain in a custom data structure. The delegate subject will subse- quently act as the delegator of this delegated role. If the delegation is *single-step*, then the delegate would not be allowed to further delegate the delegated role. However, the delegator could achieve **multi-step delegation** by setting a Boolean variable **multiStepDelegatable** to true (i.e., 1) for the corresponding delegation in the blockchain. Moreover, the delegator can control how often the delegated role can be transferred further. To this end, the delegator can specify the value of the **levelsofDelegation** variable, which is automatically set to 1 if no value is specified (default). For instance, if the delegator has specified its value equal to 5 then the delegate would be able to further delegate the role 5 times. It is important to note that another variable called **remaningLevels** tracks the number of further delegations. The **levelsofDelegation** variable remains constant during further delegations. However, the **remainingLevels** variable is automatically decremented with each possible further delegation. Consequently, our RBAC smart contract will discard further delegations when the value of the latter variable becomes equal to 0.

It is important to remark that RBAC supports two mechanisms to transfer access rights: (i) delegation of roles and (ii) delegating the individual permissions. The delegation of a role $r$ empowers the delegate to act in the delegated role $r$. This essentially means that the delegate becomes authorized for $r$ and, consequently, obtains the privileges associated with $r$. Delegating a permission $p$ gives the delegate the ability to use $p$. However, all other subjects who are associated with the delegate role also obtain the delegated rights [55]. This article focuses on both role and permission delegation operations. Both of these operations, respectively, trigger changes in subject-role relationship and role-permission relationship stored in the blockchain. Our blockchain-based RBAC supports the following transactions to create and update these delegation relationships. Table 2 shows the one-time constant resource consumption costs of these transactions.

- The *Role Assignment Transaction (RAT)* enables an administrative user to grant an appropriate role to a subject. The RAT is essentially a self-certified delegation that the *issuer* performs to grant some role to a **Subject**. RAT requires the subject' public key and the role that the issuer wants to assign. Consequently, it stores the subject-role relationship in the blockchain and shows the output when the execution is successfully completed. For the sake of simplicity, RAT grants role to a single subject **ID** only. However, the issuer can also assign the same role to multiple subjects in a single RAT transaction. It is worth noticing that only an authorized issuer can assign roles to subjects in an organization. To this end, our RBAC contract contains a special modifier that authorizes the sender of RAT. Moreover, we use a custom data structure to store the subject-role relationship in the blockchain in an auditable fashion.

- The issuer sends a **Role Update Transaction (RUT)** to modify an existing subject-role relationship stored in the blockchain. This transaction requires the public key of the EOS account of the subject and a new role to re- place the existing information in the blockchain. Consequently, it updates the subject-role relationship database stored in the blockchain and notifies the issuer about the changes that just took place. It is worth noticing that only an authorized issuer can update an existing subject-role relationship in the blockchain.

- Our proposed blockchain-based RBAC provides the **Right Transfer Transaction (RTT)** transaction to enable a subject to grant his/her available access rights to another subject. This mechanism enhances expressiveness and permits a natural administration model where the issuer can create roles and designate other (less privileged) entities to delegate these roles using **user delegation**. Consequently, a subject (delegator) who is granted a role can send a **Right Transfer Transaction (RTT)** to the blockchain to further transfer the access rights to another subject. This transaction requires the public key of the delegate subject and other relevant information such as expiry time to provide temporary delegations that are limited by time. Once the expiry time reaches, the RBAC smart contract automatically deletes the delegation from the blockchain. Moreover, it notifies the delegator that the public key of the delegate has been added to the blockchain when the execution is successfully completed. It is worth noticing that we use a custom data structure to store user delegations in the blockchain.

- The **Permission Assignment Transaction (PAT)** creates authorizations and obligations in the blockchain. Permissions in RBAC cannot be applied to subjects directly, but to a role group. This transaction allows the issuer to specify mode of the RBAC policy, name of the role, action indicating the operation to be processed by role, target object, constraint and other exceptional conditions (if applicable). The execution of this transaction creates a new role-permission relation in the blockchain as shown in Figure 2.

- Permission updating is simply the process of updating a role-permission relationship in the blockchain. Permission updating is a principal component of role maintenance in role life-cycle. The issuer sends a **Permission Up- date Transaction** to update an existing role-permission relationship in the blockchain. The issuer specifies the new permission set, constraints and exceptions to update the existing information in the blockchain.

Though our blockchain-based RBAC currently supports *passive delegation*, it could also be enhanced to support *active delegation*. In active delegation, a subject sends a request to delegate another role's permission to itself. In this case, delegator and delegate are both the same subjects. In active delegation, the role-permission relationship contains an exceptional field with A- mode. Due to the transparency property of blockchain, other subjects can check the blockchain to determine the value of this exceptional field. Therefore, the delegator would directly send a request to the blockchain for the delegation of a particular role if such a role meets the above exceptional requirements. On the contrary, passive delegation allows a subject to send delegation request to another subject. In this case, delegator and delegate are not the same subjects [21]. Consequently, the delegator sends a delegation request message to the blockchain to delegate his role permission to the delegate subject.

## 4.2 Revocation

Revocation is the reverse process of delegation that removes or retracts delgations from the blockchain. Our approach allows the delegator to remove delegations at any time. The following subsections discuss the types of revocations and address some of the issues that might introduce complexity.

Our proposed system supports temporary delegation that automatically expires with time. We call the delegation period as duration of delegation, which must be chosen carefully. The delegators should not overestimate the duration of delegation because it allows the delegate to continue to use the permissions associated with the delegated role. Moreover, the duration of delegation should not be short enough that would prevent the delegate from completing the assigned task.

Our proposed RBAC supports two types of human revocation operations (i) administrative revocation and (ii) user revocation. An administrative revocation operation requires the service of an administrative human (e.g., issuer) to revoke a delegation from the blockchain whereas a user revocation operation allows the delegator subject to remove the right transfer from the blockchain. The latter approach empowers the delegator to control the behavior of the temporary delegate. Hence, the delegator is primarily responsible to protect the system resources from the delegate. For instance, the delegator should immediately remove the delegation from the blockchain if the delegate acts badly in the delegated role. To overcome this situation, any of the original role members should be able to revoke such delegations from the blockchain. Alternatively, the issuer can also revoke the delegating role if the delegate behaves badly. In particular, our RBAC smart contract supports the following essential human revocation operations. Table 2 shows the one-time constant resource consumption costs of these transactions.

- The issuer sends a *Role Revocation Transaction (RRT)* to delete an existing subject-role relation from the blockchain. This transaction requires the subject' public key as input parameter and, consequently, it deletes the corresponding subject-role relation from the blockchain when the execution is successfully completed. Moreover, the RBAC smart contract notifies the issuer about the changes in the subject-role relationship database stored in the blockchain. It is worth noticing that only an authorized issuer can revoke an existing subject-role relationship from the blockchain. To this end, our RBAC contract contains a special modifier to authorize the sender of RRT transaction.
- The *Permission Revocation Transaction (PRT)* deletes an existing role-permission relationship from the blockchain. To accomplish this goal, the issuer specifies the identifying information of the role-permission relationship to delete it from the chain. It is worth noticing that the issuer creates role-permission relationship for each distinct role in the system. Consequently, the issuer specifies the existing role in a Permission Revocation Transaction to delete the associated role-permission relationship from the blockchain.
- Remove Right Transfer: Subjects who have further delegated access rights send *Remove Right Transfer Transaction (RTT)* to the blockchain to remove thrid-party delegations from the blockchain. This transaction is essentially the reverse operation of the right transfer transaction discussed earlier. Subjects should not only be able to transfer access rights, but they should also be capable to delete right transfers from the blockchain to protect resource misuse.

With respect to the revocation operation, a distinction should be made between strong and weak revocation. If a subject is simultaneously assigned to a senior role R1 and a junior role R2, then a weak revocation of junior R2 would only remove assignment in the junior role. However, a strong revocation of junior R2 would remove assignment in the junior role as well as assignment in the senior role R1. This is due to the fact that senior role R1 inherits the permissions from junior role R2. Indeed, a weak revocation operation can be easily accomplished because it only requires the same parameters as specified in the RAT transaction. However, strong revocation operation cannot be easily accomplished because it needs information about explicit and implicit role membership. In this example, the user would be explicitly assigned R1 and implicitly assigned R2. In other words, a strong revocation operation requires checking the role hierarchy to see whether the subject is assigned to a senior role. If such assignment exists, then it needs to be removed as well. Further details on how to explicitly and implicitly assign roles are outside the scope of this article.

Our approach supports delegator-dependent revocation such that only the delegator is allowed to revoke the delegated role. Indeed, this approach offers various advantages such as controlling the process of revocation and resolving conflict between the original role members. However, it has also some limitations such as keeping track of the delegators, the possibility for the delegate to misbehave for a long time because other members cannot revoke the delegation and dealing with the delegated role when the delegator is removed from the role membership. On the contrary, delegator-independant revocation could also be achieved by allowing any original role members to revoke the delegated role. This approach would empower the original role members to control the usage of the delegated role, which is an advantage because the delegate may act badly in the delegate role. However, it can also create conflict between the original role members. This happens because any of the original role members can delete such delegations from the blockchain.

Finally, we consider the case of cascading revocation. Suppose that Alice delegates (using user delegation) her role membership to Bob to perform some task. Alice can latter delete Bob's delegated access rights from the blockchain by sending a remove right transfer transaction to the blockchain. In this case, suppose that the issuer sends a transaction to the blockchain to remove Alice subject-role relationship from the blockchain. This will result in the automatic revocation of Bob's delegated access rights in role R (and from any roles junior to R) as well. The issuer has full authority to remove third-party delegations from the blockchain.

### 4.3 Constraints

In recent years, highly flexible devices have emerged that are daily used in any part of human life. Moreover, the widespread deployment of Internet and net- working technologies demand the protection of sensitive information that can be accessed by subjects and/or machines. The increasing usage of mobile de- vices in organizations also presents a challenge because users require access to protected resources from a variety of settings. One of the important aspects of access control is the consideration of location and time to control access to critical resources. Such constraints are important for controlling location and time-sensitive activities that are necessary in various organizations. Indeed, this situation demands the provision of customized, decentralized and context-aware role-based access control policies. Our fine-grained access control solution ef- ficiently solves this problem because the role-permission relationships support context information in access control decisions. Our approach allows the issuer to rapidly update permissions to deal with a dynamically changing contextual information. Moreover, our approach creates ***conditional permissions*** such that each permission could be associated with one or more context constraint.

The concept of constraints emerged since some of the initial RBAC systems were proposed [19] [51]. In these earliest systems, the concept of constraints was mainly focused on the restriction of the assignment of subjects and permissions to roles to enforce common access control policies, such as separation of duty. However, constraints models were not a part of these initial RBAC systems. RBAC constraints can be broadly classified into ***static*** and ***dynamic*** [56] as discussed below;

Static constraints refer to constraints that can be evaluated directly at design time or upon assignment of a role within an RBAC system (e.g. Static Sepa- ration of Duty (SoD)). Static constraints are also called ***assignment constraints*** because they control the assignment of roles and permissions. RBAC delegations essentially create a direct relation between subjects and access rights – subject S obtains the privileges associated with R only if R has been delegated to S. Consequently, constraints that restrict delegation must directly restrict this new relation between subject and access rights. Predicates provide a general approach to express RBAC constraints. We can simply use the predicate *right* to show this new relation between subject and access right R.

$$\bot \leftarrow right(S,R), \gamma \quad (1)$$

Where $\gamma$ represents any coexistence of the RBAC basic predicates (i.e., between subjects and roles, hold between roles and rights) or any other predi- cate related to subjects, roles and rights. The above constraint states that both *right(S, R)* and $\gamma$ cannot be true at the same time. For instance, we assume that the declaration *junior(X, Y)* states that subject X is junior to subject Y in the organization's role structure. Then we can use the following generic constraint to indicate that a junior cannot receive a right R that is stronger than the one that his senior owns:

$$\bot \leftarrow right(S, p); junior(S; boss); play(boss; role); hold(role; q); imply(p; q) \quad (2)$$

The constraint that a subject S who can exercise/play the role of a watchman does not possess the rights to delegate the duty of locking doors can be expressed as follows:

$$\bot \leftarrow play(S, watchman), right(S; d(lock-door, 0)) \quad (3)$$

Finally, a **separation of duty constraint** to express the idea that a Subject S who can perform task1 cannot also perform task2 can be simply expressed as follows:

$$\bot \leftarrow right(S; task1), right(S; task2) \quad (4)$$

Static constraints are enforced when tuples are added or removed. Tuples represents new instances of the subject-role relation and role-permission relation. The static constraint must be checked after a transaction insert new instances of these relations. Indeed, the result of checking static constraint would be the presence or absence of security violations. A practical implementation of the above constraints should not only verify that a security violation has occurred, but also report which instantiation of the variables *subject* and *role* causes the security violation.

On the contrary, context constraints refer to constraints that do not relate to the core elements of RBAC but are usually defined as extra conditions for access control decisions [3]. These constraints evaluate the subjects' attributes at run- time to see whether they match with the pre-defined values (e.g., context constraints such as time and location). As shown in Figure 2, context constraints are linked with RBAC permissions because RBAC authorization decisions take the role-permission relationship into consideration. A context constraint states that certain *context attributes* must match certain conditions to allow a specific operation. A *context attribute* indicates a certain environmental property whose actual value change dynamically. For instance, time, location, date, birthday, affiliation to an organization, employee's salary or even nationality can be used as context attributes. Context constraints essentially define conditional permissions such that each permission is associated with one or more context constraint. Consequently, a subject can perform an operation on an object if and only if each corresponding context constraint evaluates to true. Our RBAC smart contract contains constraints variables to store context attributes with different data types (e.g., string, date, integer, etc.). Moreover, the RBAC smart contract contains a Boolean *check constraint* function to compare the current value of a context attribute with that specified in the RBAC policy. This function checks that context constraints associated with a particular permission are not violated.

A scenario-driven role-engineering process is presented in [3] for the specification and elicitation of context constraint. This process considers the usability of a system to derive permissions and to define tasks. Broadly speaking, a scenario describes an action or a sequence of events. Thus, each scenario may consist of a number of steps and a subject must possess all permissions that are required to successfully perform the various steps of this scenario. To obtain context constraints, we first identify the goals and obstacles of the scenario under con- sideration. We then examine each goal to derive the context attributes that are required to perform this specific goal (e.g., date, time, location and so on). Ob- stacles are useful to derive context conditions because they essentially define what should not happen. Moreover, this study provides an online examination scenario where a student must possess all permissions to perform the various steps of this scenario such as fetching and editing exam documents. When we examine the goals of this scenario, we can clearly identify context attributes such as examination date, current time, exam document number and so on. Further- more, context conditions can be identified from the obstacles of this scenario such as checking to see whether the client (student) IP address is a registered one, etc. It is worth noticing that this role-engineering process is applicable to a wide range of scenarios.

Organizations collaborate to share data and functionality with collaborative partners, while controlling access to resources from unauthorized users. Indeed, access control policies play an important role to accomplish the tasks of col- laborating organizations. However, such collaborative environments require the consideration of policies that are capable to explicitly specify which subjects from which organizations (contexts) can perform operations on resources. It is worth noticing that the basic RBAC [19] does not include any mechanism to specify such kind of constraints. On the contrary, the modern RBAC supports context attributes such as affiliation to an organization, which can be used to meet the security requirements of such collaborative environments. Indeed, dynamic up- dating of large number of contexts may bring burden to the administrative user because the access control policies need to be modified when adding, removing or updating context attributes. However, our approach allows the administrative user to modify these policies when adding or removing contexts, thereby reducing the administrative burden.

Although, the significance of RBAC constraints has been realized for a long time and different techniques have been proposed to specify and enforce authorization constraints, still some constraint-related challenges have not received much attention from the research community. Initial work on constraints was mainly focused on constraints specification rather than enforcement. Authorization constraints can also be expressed in natural languages, such as English, or in more formal languages. Sandhu et. al [57] have proposed a role-based constraints specification language (RCL) which can be used to express authorization constraints. RCL can be used by security researchers who are mainly responsible to understand the objectives of organizations and express security decisions supporting these objectives. On the contrary, the administrative user is mainly responsible to process day-to-day operations. Another advantage of RCL is that it provides greater flexibility, sharing common semantics to express constraints. There are too many possibilities and variations of RBAC constraints. RCL is a powerful language because it has the potential to express many constraints previously identified in the literature. Humans prefer natural language specification due to ease of understanding, but such specification may be prone to ambiguities. Furthermore, initial research work mainly focused on separation of duty. However, other types of constraints such as Binding of Duty (BoD) has received less attention from the research community. A simple example of BoD would be that a subject can be assigned to role X only if he/she is already assigned to role Y. To the best of our knowledge, no research work is dedicated to this topic.

Constraints can be put into system design in two different ways. First, authorization constraints can be specified in predicate logic (PL). Context constraints of RBAC are defined through the formula construct which can be specified in ambient logic (AL). Together these constructs provide a powerful technique to specify RBAC policies for smart contracts. Alternatively, constraints specification can be injected into a unified modelling language (UML) representation of RBAC as realized by [58]. UML together with Object Constraint Language (OCL) have also been recently considered as fit for blockchain ontology [59]. OCL is a declarative language which can be used to describes the rules related to UML and is part of the UML standard. OCL is a precise text language that can be used with any metal-model to express constraints.

## 5 Advantages

Our proposed blockchain-based RBAC provides various advantages because the blockchain performs the functions of policy management and evaluation. Therefore, our proposed system inherits the advantages of blockchain technology such as auditability, privacy, distributed with no single point of failure, immutability, availability and so on. This section discusses the advantages of our proposed approach.

## 5.1 Privacy

Current blockchain systems provide the auditability property to ensure that all the users' actions are transparently stored in the blockchain. Consequently, auditability may lead to privacy problems. Our blockchain-based RBAC stores all the access control policies, and access decisions in the EOS blockchain. However, public blockchain platforms are more vulnerable to privacy problems because the platform is open to everyone. Therefore, users and even miners can read all the data published in the blockchain. To tackle this issue, current blockchain platforms offer the pseudonymity feature, which ensures that registered users are associated with unique randomly generated identifiers that show no information about the real identities of users. Thanks to the pseudonymity feature, all the role-based access control policies and access requests, although transparently stored, but show no information about the real identities of the users who conducted them because we store all the information with anonymous IDs. However, the issuers and the resource owners need to know the real identities of users when assigning roles to users. Therefore, an attacker is not capable to learn sensitive information from the blockchain since anonymous IDs are stored there. Finally, an attacker must forge a digital signature or control the majority of computing power in the network to alter access permissions in the blockchain. Blockchain platforms exploit a signature-based security to prevent the former and a decentralized consensus mechanism to prevent the latter.

Furthermore, some techniques called deanonymization attacks have been proposed to break users' pseudonymity. However, these techniques mainly focus on the Bitcoin blockchain platform and are based on graph-based analysis. Furthermore, researchers have recently investigated that graph-based analysis can be conducted on the Ethereum platform [60]. It is worth noticing that similar techniques are possible for new blockchain platforms. Therefore, new systematic approaches are necessary to measure and identify the possibility of such attacks on new blockchain platforms such as EOS and Steller. Moreover, a number of studies have analysed the vulnerabilities in Ethereum smart contract and different attacks have been identified such as the re-entrancy attack. Therefore, our research work also opens the door to investigate security vulnerabilities in EOS smart contracts.

Alternative proposals that offer enhanced privacy protection have recently emerged such as CoinJoin. CoinJoin allows users to route their transactions through a centralized mixing service, which hides the relationships between the originators and the receivers of transactions. However, such techniques largely rely on the centralized mixing service, which becomes a single point of failure and trust. Consequently, the mixing service may steal assets that are routed through it. To overcome the limitations of CoinJoin, a number of alternative proposals have emerged such as Mixcoin, which aims to eliminate the risk of theft by making the mixing service responsible if it steals the assets of a user. However, sophisticated cryptographic techniques are necessary to ensure complete information disclosure. For instance, zero-knowledge proofs (ZKP) can be used to hide the private information of users. However, zero-knowledge proofs are complex and have high computation and storage cost [61].

Finally, built-in cryptographic primitives can be used to implement privacy-preserving smart contracts. The EOS blockchain platform provides built-in cryptographic primitives such as SHA-3 to encrypt the contract data and calls. EOS SHA-3 functions are based on an instance of the keccak algorithm. Therefore, the blockchain will store obfuscated information once we take cryptographic hash of the contract's data [16]. The Ethereum platform also provides built-in cryptographic primitives to implement privacy-preserving smart contracts. However, the Ethereum platform incurs extra gas cost for each cryptographic operation defined in the smart contract.

## 5.2 Auditability

The role-based access control policies comprising subject-role relationship and role-permission relationship are transparently stored in the blockchain. Storing these relationships in the blockchain provides various advantages such as the prevention of false denial and the transparent verification of permissions. If the service provider or third parties falsely deny an operation on a certain resource to a subject, that subject can check the blockchain at any time to see whether he has been assigned an appropriate role. Moreover, subjects can check the role- permission relationship in the blockchain to see what permissions have been associated with the roles that are assigned to them. Therefore, subjects can take the necessary actions on time such as informing the organization (or the issuer) if some required permissions are not present in the blockchain. Due to the immutability property of blockchain, all the actions performed on a smart contract are transparently stored in the log, which is accessible on the blockchain. Hence, blockchain provides long term auditability.

The blockchain provides a trusted execution environment for performing access control decisions and provides a trusted storage for storing access control policies. On the contrary, if the access control policies are stored in a medium other than blockchain, then it is possible that an adversary could alter the access control decision process or store fake access decision information in the associated log file. Obviously, the adversary cannot perform such attacks on the blockchain unless it controls more than half of the computing power in the blockchain network, which is assumed to be hard. Our approach improves trust through decentralization and by storing the organizations policies on the blockchain. It is worth noticing that adopting the Ethereum blockchain for enforcing role-based access control policies introduces some costs to perform policy ownership transactions such as role/permission assignment, update and revocation. A description of the gas costs for our Ethereum-based implementation is given in [62]. Consequently, our research work adopts the EOS permissioned blockchain platform to reduce the deployment and execution costs. Indeed, the EOS blockchain provides a trusted storage and execution environment, allowing only trusted block producers to append new blocks to the chain, and supports gasless transactions to eliminate fees associated with essential transactions such as contract deployment, policy creation, update and revocation.

It is important to remark that auditability in our proposed system concerns the storage of role-based access control policies and the evaluation of these policies, i.e., the service provider or third parties check access rights by sending the traditional access control triplets (subject ID, operation and object) to our RBAC smart contract. The smart contract first examines the subject-role relationship to fetch the assigned role and it then uses this information to check the permission associated with the assigned role. Consequently, our RBAC smart contract stores audit data in the associated log file, which can be analysed latter to discover or diagnose security violations. Indeed, audit data demands protection from modification by an attacker. Thanks to the tamper-resistant characteristic of blockchain, an attacker cannot modify the audit data stored in the blockchain.

### 5.3 Availability, Integrity and Confidentiality

Availability and integrity result from the persistency property of blockchain. That is, once some data is added to the blockchain as a part of a valid block, it always becomes available and impossible to modify or delete it (integrity). Therefore, our blockchain-based RBAC provides a trusted and always available environment that permits service providers or third parties to have tamper-proof evidence of the users' access rights before allowing them to perform operations on resources. As such, our blockchain-based RBAC successfully defeats malicious users who may try to create fake policies in the blockchain. It is important to remark that each organization has a unique RBAC smart contract deployed on the blockchain such that only the administrative user (issuer) is capable to perform policy-ownership operations such as policy creation, update and revocation. EOS smart contract contains special modifier to authorize the sender of policy-ownership transactions. Therefore, our RBAC smart contract automatically rejects requests that are sent from non-administrative users. Moreover, the blockchain infrastructure is more resilient to availability attacks such as Denial of Service (DoS) attacks and impersonation attack. On the contrary, existing solutions that are based on a central server are vulnerable to the single point of failure and DoS attacks. For instance, the rule-based access control system [63] is proposed where policies are expressed in terms of relationships existing between subjects. However, the access control rules, decision module and certificates are all stored and managed in a central node, which may become a single point of failure. Finally, confidentiality checks if the information is leaked to unauthorized peers. It is worth noticing that blockchain utilizes public/private key pair as an addressing mechanism. Therefore, our blockchain-based RBAC system has a built-in confidentiality and authorization feature since the issuer private key is used to sign each policy ownership transaction such as policy creation, update and revocation.

## 6 Performance Evaluation

Every transaction sent to execute the code of a smart contract in the EOS blockchain system requires the consumption of resources [64]. The EOS blockchain classifies resources into three types: computing power, storage and network bandwidth. For the sake of simplicity, we simply indicate them as CPU, RAM and NET, respectively. A transaction sent to execute the code of a smart contract can only be successful if the sender possesses enough resources. Otherwise, EOS asks the owner to acquire resources by spending EOS tokens. An EOS token is a utility token, and owning the tokens (staking) provide bandwidth and storage on the EOS blockchain. It is worth noticing that registered users freely obtain tokens from the EOS blockchain network [12]. Moreover, EOS provides an essential characteristic to delegate resources. Token holders delegate computation and bandwidth to other users in the network. This essentially means that if token holders are not utilizing their allocated CPU and NET, then they can rent or delegate these additional resources to other users in the EOS system. However, unlike bandwidth and computation, token holders cannot delegate RAM [16]. Unlike the Ethereum platform, EOS smart contracts and DAPPs work without transaction fee. Users with certain number of tokens obtain a percentage of the EOS computational power. Moreover, the EOS platform provides free usage to the registered users. However, users only need EOS tokens to send transactions to perform operations on the blockchain [65] [34]. Finally, EOS allocates the contract owners the resources of Block Producers (BPs) when they spend their tokens to acquire RAM. Indeed, this technique aims to efficiently manage the resources of Block Producers. More importantly, EOS distributes resources proportionally to the number of EOS tokens staked by a user. For instance, if an EOS user stakes 1% of the total EOS tokens, then that user can utilize 1% of the system resources as well. Moreover, EOS releases the allocated CPU and NET when the user unstake tokens [16].

### 6.1 Experiment Setting

We use the C++ programming language to write the code of our RBAC smart contract, deploy and test it on the EOS Kylin Testnet. Users interested to write smart contracts for the EOS blockchain need to install the Contract Development Toolkit (CDT). The CDT tool contains the *eosio_cpp* command that compiles the contract's C++ file to Web Assembly (WASM) format. This is how it generates the related Application Binary Interface (ABI) files to enable user action to convert between JavaScript Object Notation (JSON) and their binary form [16]. Unlike the Ethereum platform which customizes a virtual machine for executing code, EOS chooses WASM so that developers can write smart contracts in different programming languages (e.g., EOS currently supports the C++ programming language). Moreover, EOS provides local testnets as well as mainnets[4] to compile and deploy smart contracts. We installed Ubuntu 18.04

---
[4] Kylin and Jungle testnets are currently available for testing EOS DAPPs and smart contracts

on a laptop computer containing core i7 CPU, 16 GB RAM, 500 GB SSD and installed the EOSIO version 1.8.1. We use the methods described in [16] to built a local testnet for the purpose of testing and evaluation. To further evaluate the performance of our proposed RBAC contract, we deploy our RBAC smart contract on the EOS Kylin[5] testnet.

### 6.2 Results

Table 2 shows the one-time constant resource consumption costs of the different functions of our proposed RBAC smart contract. Our experimental results reveal that the *Permission Assignment Transaction (PAT)* requires higher resource consumption cost than the remaining functions of the smart contract. This is due to the fact that PAT requires multiple elements (e.g., role, resource, action, constraint) as input parameters that needs more CPU and Net usage compared to the costs of the other functions. The service providers or third-parties use the *Check Access* function of our RBAC smart contract to determine access permission in the RBAC policy database. This function first checks the subject- role relationship in the blockchain to verify the role membership of the requesting subject. It then checks the associated role-permission relationship to determine access permission in the blockchain.

Table 2: CPU and Net costs of the RBAC contract

| RBAC Transaction | CPU Usage($\mu$s) | Net Usage(Bytes) |
|---|---|---|
| Role Assignment | 606 | 168 |
| Role Update | 347 | 168 |
| Role Revocation | 209 | 104 |
| Check Access | 305 | 104 |
| Right Transfer | 511 | 176 |
| Remove Right Transfer | 254 | 104 |
| Permission Assignment | 856 | 160 |
| Permission Update | 570 | 160 |
| Permission Revocation | 230 | 104 |

**Network Latency** Figure 5 shows the network latency experienced and compares it with the latency of the RBAC system on the Ethereum blockchain network. We setup our own EOS private testnet to measure the latency of our proposed RBAC system. An EOS local private testnet can be configured with different number of miners who process all the transactions to the smart contract and shows the latency (in milliseconds). Our experimental results clearly show that RBAC on the EOS blockchain requires on average 46.23 ms latency. However, the RBAC system on the Ethereum blockchain requires approximately 225 ms average network latency [66]. It is worth noticing that the service providers may exploit a local cache to fetch the policy database from the smart contract for authorization validation. In this case, the network latency will further decrease.

---
[5] https://www.cryptokylin.io/

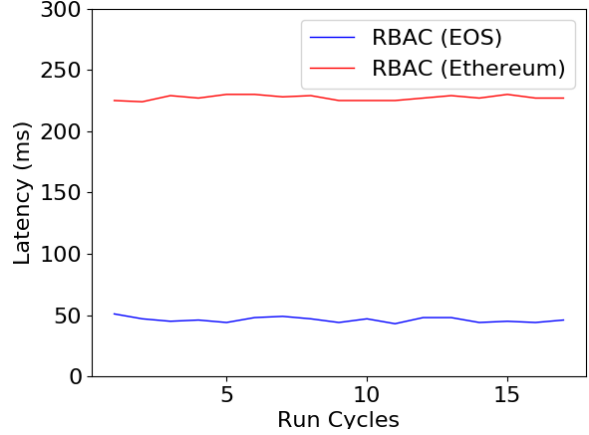

Fig. 5: Network Latency Comparison between Ethereum and EOS

**Block Generation Time (BGT)** Block generation time defines the average amount of time to mine the next block in the blockchain network. A powerful feature of the EOS blockchain is that it creates blocks every 0.5 seconds. A Block Producer (similar to a miner) node needs some waiting time when it completes a given transaction before 0.5 seconds. It then broadcasts the processing block to other block producers in the network. Consequently, users experience a faster chain experience resembling a web app or a payment card. We can use the following equation to calculate the time from when the transaction is first processed to when the block containing that transaction is confirmed in the blockchain network [67].

$$BttT_u = \frac{\Sigma_{Tx}(t_{TxConfirmed} - MAX_{block}(t_{TxEXEDONE}))}{Count(Tx\ in\ (t_i, t_j))}(s/tx) \quad (5)$$

where $t_{TxConfirmed}$ indicates the time when the transaction is first confirmed and $MAX_{block}(t_{TxEXEDONE})$ indicates the timestamp when the execution of the final transaction in the same block containing Tx finishes. We can also use the following equation to find the average block generation time (BGT) to know the whole system.

$$BttT = \frac{\Sigma_u(BttT_u)}{N}(s/tx) \quad (6)$$

The role-based access control policies can be valid only if miners have confirmed the policy-related transactions to the RBAC smart contract and these transactions are added to new blocks in the network. Both Bitcoin and Ethereum assemble all broadcasted transactions into blocks. The Proof-of-Work (PoW) mining algorithm is used where every node computes a puzzle. For Ethereum, the block generation time is directly proportional to the transaction rate. The Ethereum implementation has significantly reduced the block generation time from 10 min to 15 seconds. However, the delegated proof-of-stake (DPOS) mechanism is more robust since witnesses are voted by the stakeholders, leading to lower block generation time and shorter time to confirm transactions. Figure 6 shows a comparison of the block generation time between the two popular smart contract platforms. For Ethereum private testnet, we initially conducted an experiment by configuring one miner to mine blocks. The block generation time decreases and becomes stable when the miner count reaches 4 [41]. However, the EOS blockchain produces new blocks every 500 milliseconds irrespective of the number of miner. The given results clearly reveal that the EOS blockchain is the future of decentralized applications to achieve a wide range of applications including efficient access control solutions for a wide range of domains.

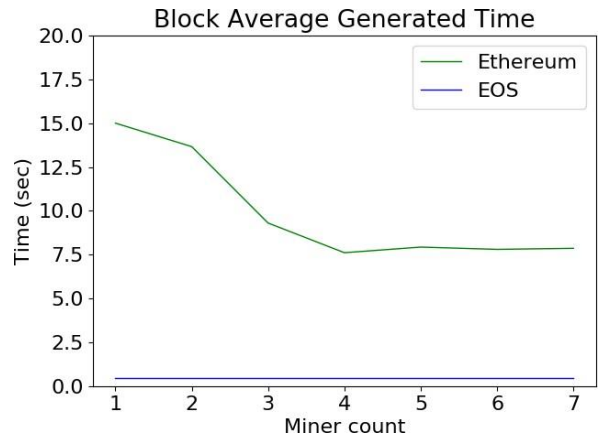

Fig. 6: Comparison of Block Generation Time

**Contract Execution Time** Contract execution time can be defined as the total amount of time (in seconds) that the blockchain platform takes to execute and confirm all transactions to the smart contract. In blockchain systems, the execution time depends on the program logic implemented by the smart contracts. We can use the following equation to determine how transactions are executed for a peer $u$ in the blockchain network [67]. For the sake of simplicity, we abbreviate transaction as Tx, as shown below:

$$CET_u = \frac{\Sigma_{Tx}(t_{TxEXEDONE} - t_{TxEXESTART})}{Count(Tx\ in\ (t_i,\ t_j))}(s/tx) \qquad (7)$$

where $t_{TxEXESTART}$ indicates the time when the transaction execution starts and $t_{TxEXEDONE}$ indicates the time when the execution finishes. We can use the following equation to take the average:

$$CET = \frac{\Sigma_u(ET_u)}{N}(s/tx) \qquad (8)$$

Figure 7 shows the average execution time for the two popular smart contract platforms for different transaction volumes. The average is taken by repeating the experiment 20 times for each transaction volume. The results reveal the difference between the performance of EOS and Ethereum blockchains when the number of transactions grows from 5000 to 10000. It is worth noticing that a large number of transactions are broadcasted to the public blockchain network. Therefore, we also investigate the performance of both platforms under high transaction volume. The given results reveal that the execution time for both blockchain platforms increases as we increase the number of broadcasted transactions. In particular, the execution time of the EOS blockchain is 42 times less than that of Ethereum at low transaction rate. However, the difference between the execution time decreases as the transaction rate further increases. Consequently, EOS outperforms Ethereum under both low and high transaction rates, thereby making it a suitable platform to design access control solutions for a wide range of domains.

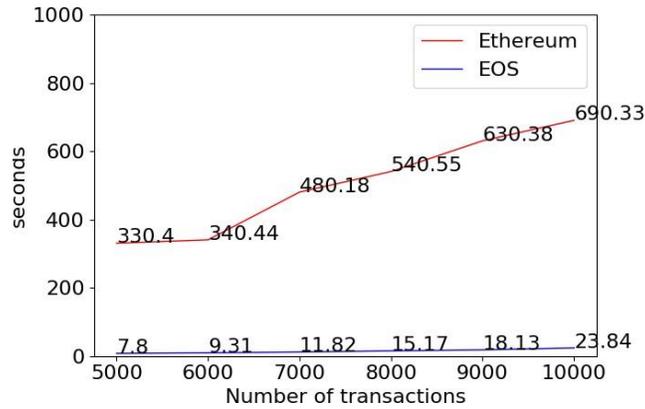

Fig. 7: Execution time comparison between Ethereum and EOS

**Transactions Per Second** Unlike the Ethereum platform, the EOS blockchain shows better performance for deploying and executing the code of smart contracts. Throughput can be simply defined as the number of transactions per second, which is measured in a period of time [67]. The following equation shows how to compute the throughput of a peer $u$ during a time interval ranging from $t_i$ to $t_j$.

$$Throughput_u = \frac{Count(Tx\ in\ (t_i, t_j))}{t_j - t_i}(txs/s) \qquad (9)$$

When considering the throughput of N peers, we can use the following equation to take the average:

$$Throughput = \frac{\sum_u (Throughput_u)}{N}(txs/s) \qquad (10)$$

Figure 8 shows throughput comparison among the popular blockchain platforms. Earlier PoW-based blockchain platforms such as Ethereum and Bitcoin require a large amount of computing power for hash calculation, which results in a very low throughput. Thanks to the advancement in consensus algorithms [14], the EOS blockchain offers improved throughput and processing abilities. The EOS blockchain platform uses an efficient DPOS consensus algorithm, which remove meaningless computing power loss and lead to higher throughput.

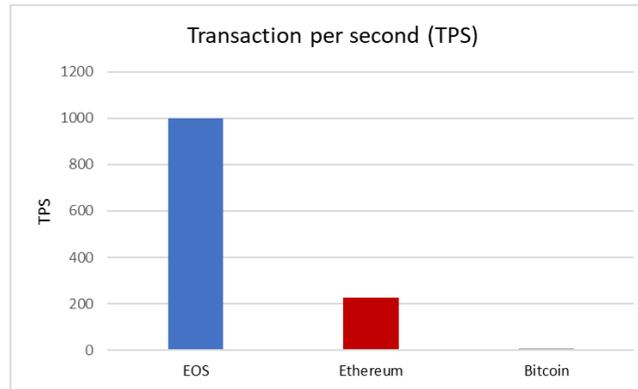

Fig. 8: Throughput Comparison among the popular blockchain platforms

**Resource Consumption** This experiment aims to evaluate the effects of the number of matches in the RBAC policy on resource consumption in the EOS blockchain network. A match in the RBAC policy database essentially consists of validating the granted role in the subject-role relationship database and checking the associated role-permission relationship in the blockchain to see whether the requesting subject can perform the requested operation on a certain resource. Consequently, our RBAC smart contract returns true when a match is found, otherwise, it returns false. In other words, we wanted to test the output of the *Check Access* function to see how does it behave when we simultaneously call it with different subject IDs. To accomplish this goal, we randomly change the IDs of the subjects when determining the access permissions in the blockchain. Figure 9 shows the average resource consumption in the EOS blockchain network when the *Check Access* function executes with different subject IDs.

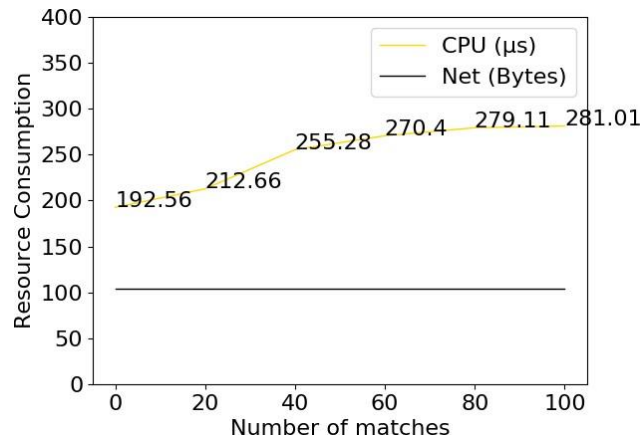

Fig. 9: Number of matches vs. Resource Consumption

## 7  Conclusion and Future Work

This research work proposes a scalable role-based access control system for both intra-organizational and inter-organizational access control. The administrative user (issuer) in an organization assign permissions to roles and grants suitable roles to subjects according to the organization's requirements. Both subject-role relationship and role-permission relationships are transparently stored in the blockchain. When a subject sends access request to the service provider(s) to perform an operation on a certain resource, the service provider(s) references these relationships in the blockchain to check the status of access permission. Therefore, our decentralized blockchain-based RBAC efficiently solves the access control problem in both scenarios. Moreover, we propose a scalable RBAC framework to allow resource owners to control access to their sensitive contents. Each resource owner sends transaction to the RBAC smart contract to create role-permission relationships. The resource owner can also delete these relationships at any given time. Currently, the General Data Protection Regu- lation (GDPR) protects personal information and it requires resource owners to acquire a user consent to store and manage his/her personal information. The GDPR also ensures " the right to erasure" i.e. the user has the right to delete personal information at any given time. In any blockchain based solution this right can be satisfied by encrypting personal information and by destroying the decryption key to prevent further accesses to an erased information.